\documentclass{article}
\usepackage{PRIMEarxiv}

\usepackage[utf8]{inputenc} 
\usepackage[T1]{fontenc}    
\usepackage{hyperref}       
\usepackage{url}            
\usepackage{booktabs}       
\usepackage{amsfonts}       
\usepackage{nicefrac}       
\usepackage{microtype}      
\usepackage{lipsum}
\usepackage{graphicx}

\usepackage{algorithm}
\usepackage{algpseudocode}
\usepackage{colortbl}
\usepackage[dvipsnames,table,xcdraw]{xcolor}
\usepackage{hyperref}
\usepackage{amsmath}
\usepackage{amssymb}
\usepackage{braket}
\usepackage{placeins}
\usepackage{makecell}
\usepackage{multirow}
\usepackage{caption}
\usepackage{comment}
\usepackage{enumerate}

\usepackage{tikz}
\newcommand{\ballnumber}[1]{\tikz[baseline=(myanchor.base)] \node[circle,fill=.,inner sep=1pt] (myanchor) {\color{-.}\bfseries\footnotesize #1};}

\usepackage{booktabs}

\newif\ifARXIV
\newif\ifIOP
\ARXIVtrue
\IOPfalse 

\title{Data Virtualization for Machine Learning}

\author{
  Saiful Khan$^{1}$, Joyraj Chakraborty$^{2}$, Philip Beaucamp$^{2}$, Niraj Bhujel$^{1}$, Min Chen$^{2}$ \\[0.5em]
  $^{1}$Scientific Computing, Rutherford Appleton Laboratory \\
  Science and Technology Facilities Council (STFC), OX11 0QX, United Kingdom\\
  \texttt{first.last@stfc.ac.uk}  \\
  [0.5em]
  $^{2}$ Department of Engineering Science, University of Oxford, \\ 7 Keble Rd, Oxford OX1 3QG, UK \\ 
  \texttt{\{first.last\}@eng.ox.ac.uk} \\
}

\begin{document}
\maketitle

\begin{abstract}
Nowadays, machine learning (ML) teams have multiple concurrent ML workflows for different applications. Each workflow typically involves many experiments, iterations, and collaborative activities and commonly takes months and sometimes years
from initial data wrangling to model deployment. Organizationally, there is a large amount of intermediate data to be stored, processed, and maintained. \emph{Data virtualization} becomes a critical technology in an infrastructure to serve ML workflows.
In this paper, we present the design and implementation of a data virtualization service, focusing on its service architecture and service operations. The infrastructure currently supports six ML applications, each with more than one ML workflow. The data virtualization service allows the number of applications and workflows to grow in the coming years.  
\end{abstract}

\section{Introduction}
\label{sec:introduction}

Developing machine learning (ML) models is a common activity in many disciplines, ranging from the sciences to digital humanities, from sports to cybersecurity, and from business and finance to engineering. In all ML workflows for scholarly research and applications, there are numerous experiments that often take multiple months -- and sometimes a few years -- to complete. There usually exist different processes for reorganizing the original data in a variety of ways, such as data merge and integration; variable selection; label selection; data reformatting, randomization, and normalization; data partition for training, validation, and testing; and so forth. This list of \emph{data transformation} increases rapidly when one includes other more complicated data wrangling and preparation tasks, such as data cleaning, labeling, augmentation, and feature extraction. A variety of data transformation tasks result in differing datasets for storing more or less the same information. In many ML workflows, these data transformation tasks are performed by individual ML developers, leading to repetitive efforts, ad hoc data management, and unauditable data provenance. It is therefore highly desirable to introduce data virtualization to enhance services to support ML workflows, especially in organizations with many different ML activities.

\emph{Data virtualization} is not a new topic and it has been extensively studied in the field of service computing (e.g., \cite{halevy2005,gu2024}). Much of the past research effort has been focused on accessing data remotely and avoiding data duplication. While this remains an important aspect of data virtualization, the aforementioned requirements from ML workflows suggest a need for data virtualization with smaller scale virtualization operations at a much higher frequency. In addition, it is highly desirable for such virtualization operations to perform some data processing tasks in addition to data accessing, as otherwise the human effort of processing a series of virtualized datasets remains costly. When an ML development environment is supported by data virtualization, we can anticipate several benefits, including storage saving, data sharing, consistent and repeatable data processing, data provenance, and so on.

In this paper, we present the design and implementation of a data virtualization service for supporting ML developers who work on many ML workflows concurrently. In particular:
\begin{enumerate}[a.]
    \item
    We propose to extend the concept of data virtualization to address the needs for creating and managing a large number of virtual datasets in ML workflows and to enable virtual datasets to be specified with both path links to proceeding datasets as well as call links to data transformation functions.
    \item 
    We designed and implemented a data virtualization service with a service-oriented architecture that featuring a number of components that were purposely designed for creating, querying, and processing virtual datasets, including a YAML-based specification standard for virtual datasets, a data query engine, a data processing engine, among others.
    \item
    We explored ML workflows in six applications, confirming the benefits of data virtualization in ML workflows. Meanwhile, these application case studies provided testbeds for the data virtualization service as well as informed many design decisions, e.g., identifying common data transformation functions (i.e., a subset of service operations) to be supported. 
 
\end{enumerate}

 
\section{Related Work}
\label{sec:related-work} 

Data virtualization has evolved since the early days of Enterprise Information Integration (EII) systems, which were aimed at providing unified access to heterogeneous data sources without centralizing them in data warehouses. Early EII platforms~\cite{halevy2005} used mediator-wrapper architectures to map queries over virtual schemas to subqueries executed on the original sources. 
However, the limited scalability, performance, and source heterogeneity of these systems motivated the development of data federation systems. A comprehensive survey of data federation systems was presented by Gu et al.~\cite{gu2024}. The authors found that industrial virtualization required a significant system infrastructure and concludd that high performance would be contingent on an advanced architecture \cite{gu2024}.
Data federated systems utilized federated queries in various languages (e.g., SQL, SPARQL) and dynamic schema mapping to handle a wide range of data.
Gadepally et al. \cite{gadepallyBigDAWG2017} extended this work by integrating multiple data models and engines, including relational, key-value, and array databases. 

The virtual knowledge graph~\cite{xiao2019} introduced flexible graph-based models in place of rigid relational schemas.
The research outlined the VKG paradigm and put forth evidence that data integration and access could be time-consuming, inflexible, and incur high costs. Furthermore, the authors highlighted that many vendors offered limited scalability, resulting in inefficiencies. Therefore, they advocated for a more comprehensive system integration and identified that system architectures would enable improved deployment \cite{xiao2019}. In a survey on performance inference in virtualised systems, Lin et al. \cite{Lin2023} demonstrated that inference between virtual machines could adversely impact performance and operational costs. Lans \cite{lans2022} and Muniswamaiah et al. \cite{muniswamaiah2019} introduced caching, parallel query execution, and hybrid approaches that blended data virtualization with selective materialization to meet performance and scalability demands. 
Jarwar et al. \cite{jarwar2019} introduced microservice-based designs to enhance modularity and availability in federated environments.
Balakrishnan \cite{balakrishnan2024} presented a high-level design that incorporated multiple databases within a single virtualization layer to optimize big data analytics and machine learning, through there was no report about the implmentation of this design.

Kennedy et al. \cite{Kennedy2023} provided evidence for accelerating virtualization to improve performance in cloud computing. Zhang et al. \cite{Zhang2020} reported the open-source GiantVM -- which is a distributed type-II hypervisor for enabling virtualization of many-to-one combination of resources from various physical machines. 
Chatziantoniou and Kantere~\cite{Chatziantoniou2021} presented DataMingler -- a system that facilitates the implementation of data virtual machines, offering schematic flexibility, rapid deployment, and agility in data integration and handling of large datasets. 
Scheiderer et al.~\cite{Scheiderer2020} proposed a distributed simulation-as-a-service architecture that supported the training of a reinforcement learning agent using remote simulation, and provided practical evidence that virtualization depended on a facilitating architecture to be scalable. They stated that the current implementation of their architecture could be improved in terms of scalability.

Almost all previous research effort on data virtualization focuses on data access, i.e., accessing explicit data via virtual data. In ML workflows, in addition to data access, there are many data wrangling activities, which are typically unglamorous but costly. In this work, we investigate how data virtualization can address the challenges in ML workflows in general and data wrangling activities in particular.

\section{Problem}
\label{sec:Problem}
In many companies or organizations that have substantial machine learning (ML) activities, one major challenge is to provide cost-effective data services to handle a variety of data used and generated in ML workflows. Such data includes not only the source data and model testing results, but also different forms of intermediate data originating from various data transformation processes (e.g., data wrangling, data processing, feature extraction, data partitioning, and so on). Examples of some common data transformations are given below:
\begin{itemize}
    \item \textbf{Merge} --- Given $k$ datasets of a similar nature from different data sources (e.g., windfarms A, B, C), one needs a merged dataset.
    \item \textbf{Integration} --- Given $k$ datasets of different nature from different data sources (e.g., weather data, river data, etc.), one needs an \textbf{integrated} dataset.
    \item \textbf{Selection} (Variables) --- Given a multivariate dataset with $d$ variables (dimensions), one needs to focus on a subset of variables and thus a simplified dataset with $d' < d$ variables (e.g., selecting the two temperature sensors -- among 238 sensors/features -- in the Windfarm dataset C.)
    \item \textbf{Selection} (Labels) --- Given a dataset with $l$ multivariate data labels, one needs to develop a model for each label, and thus $l$ datasets (e.g., object shape labels for a shape classifier, occlusion labels for occlusion detection, and object size labels for a size classifier, and so on).
    \item \textbf{Normalization} --- Given a collection of data objects, some ML training and testing methods can only deal with data objects of a specific format (e.g., images of the same size and aspect ratio, time series data in a pre-defined data range, and so on).
    \item \textbf{Reorganization} --- Given a dataset, one needs to reorganize the data in order to use a specific type of machine learning method. For example, a long time series may be reorganized into $N$ data segments, time series prediction models are commonly trained with the former, while time series classification models are commonly trained with the latter.
    \item \textbf{Feature Extraction} --- Given a dataset, one needs to transform each data object into its feature representation (e.g., a feature vector for an image, or a feature time series for a raw time series with measured data, and so on). 
    \item \textbf{Partition} --- Given a dataset with $N$ data objects (labelled or unlabelled), partition it into $a\%, \; b\%, \; c\% \; (a,b,c>0; \; a+b+c = 100\%)$ of training, validation, and testing datasets with random object selection. Multiple experiments may involve different specifications of $a, \; b , \; c$ and random seeds.  
\end{itemize}

In abstraction, different data transformations can be denoted as different functions $F_k, k = 1, 2, \ldots, N_F$, such that
\begin{equation}\label{eq:Transformation}
     F_k: \mathbf{X}_1 \times \mathbf{X}_2 \times \ldots \times \mathbf{X}_{N_S} \longrightarrow
     \mathbf{Y}_1 \times \mathbf{Y}_2 \times \ldots \times \mathbf{Y}_{N_D}  
\end{equation}
where, each $\mathbf{X}_i$ is the set of all possible $i$-th source datasets $(i=1, 2, \ldots, N_S)$, and each $\mathbf{Y}_i$ is the set of all possible $j$-th destination datasets $(j=1, 2, \ldots, N_D)$. Here, the input datasets for and output datasets from each function $F_k$ are ordered, since this is the convention to define most data transformation functions.
 
In ML workflows, one major shortcoming of the conventional approach is that individual ML developers regularly decide how data transformations are implemented and how the output datasets are stored. Some ML developers effortlessly implement data transformation programs that generate ``explicit'' output datasets, which take up a fair amount of space. For example, given large collections of images $\mathbf{X}_\text{image}$, ML developers commonly partition them into three datasets $\mathbf{X}_\text{trn}$, $\mathbf{X}_\text{vld}$, $\mathbf{X}_\text{tst}$ for training, validation, and testing. If an ML developer wishes to experiment with several partitioning schemes, they sometimes store multiple collections of $\mathbf{X}_\text{trn}$, $\mathbf{X}_\text{vld}$, $\mathbf{X}_\text{tst}$ for various partitioning schemes (e.g., different combinations of $(a\%, b\%, c\%)$). Alternatively and more commonly, one may simply replace old partitioned datasets with new ones in iterative experimentation, leading to the loss of data provenance.

In some cases, the destination datasets may take more space than the source datasets. For example, when a long time series is transformed into $N$ data segments using a sliding window of length $W$, the size of the destination dataset (i.e., a collection of time series segments) could be close to $W$ times the original time series (i.e., a long time series). Hence, many ML developers avoid this type of data transformation, although having a collection of $N$ data segments would allow ML developers to make use of more ML techniques.

Although some ML developers write more sophisticated programs to be used in ML workflows by passing path links to training, validation, and testing programs, the lack of a consistent and coherent approach within each workflow and across different workflows undermines quality management in engineering ML models. In Table \ref{tab:Comparison}, some major shortcomings of a conventional ML model development environment are shown in the second column (without data virtualization). In the next section, we discuss how data virtualization can address these shortcomings.

\begin{table}[h!]
\scriptsize
\centering
\caption{Comparison of types of model-development environments: individually-managed data transformations without data virtualization services (middle column) vs. infrastructure with data virtualization services (right column).}
\label{tab:Comparison}
\vspace{1mm}
\begin{tabular}{%
    @{}p{2cm}%
    @{\hspace{2mm}}p{5cm}%
    @{\hspace{2mm}}p{5cm}@{}%
    }
    
    \toprule
    \textbf{Category} & \textbf{Without Data Virtualization} & \textbf{With Data Virtualization} \\
    \midrule
    

      \raggedright\textbf{A$_1$ Version control \& reproducibility} &
      Inconsistent data transformations practices, difficulty reproducing experiments due to missing data versions, or transformation details. &
      Systematic versioning of datasets, transformations, and configurations with easy reproducibility. \\
    \midrule
    
      \raggedright\textbf{A$_2$ Data operation consistency} &
      Inconsistent data operations (e.g., selection, feature extraction, partition, etc.) lead to issues in comparability of trained models. &
      Consistent data operations, potentially through standardized APIs and UIs with validated transformation pipelines. \\
    \midrule
    
      \raggedright\textbf{A$_3$ Data provenance} &
      Relying on individuals to maintain data provenance, demanding significant amount of effort from individuals, commonly leading to very limited data provenance. Datasets for experiments are often deleted after experiments. &
      Automated maintenance of data provenance with complete lineage tracking from source to results. \\
    \midrule

      \raggedright\textbf{A$_4$ Experiment management} &
      Manual experiment tracking, difficulty comparing results across different data preprocessing approaches, informal parameter management leading to irreproducible results. &
      Systematic experiment tracking with automated model parameter and metrics logging, comparative analysis tools, and reproducible experiments. \\
    \midrule
    
      \raggedright\textbf{A$_5$ Integration with ML workflow} &
      Manual integration between data preparation, model training, validation, and deployment phases, leading to workflow fragmentation and inefficiencies. &
      Seamless integration across the entire ML lifecycle with automated pipeline orchestration and standardized interfaces between phases. \\
    \midrule
    
    
     \raggedright\textbf{B$_1$ Storage Management} & Many ML developers commonly create physical copies of data when only a small portion is needed, or links to source data may not be easy to maintain. This leads to significant space waste. & Infrastructure enables reliable virtualization with logical data access, eliminating unnecessary physical copies. \\
    \midrule
    
     \raggedright\textbf{B$_2$ Scalability} & Limited by individual machine capabilities, manual scaling requires significant developer effort, performance bottlenecks difficult to identify and resolve. & Automatic scaling across distributed resources, intelligent workload distribution, and performance monitoring with optimization recommendations. \\
    \midrule
    
     \raggedright\textbf{B$_3$ Data discovery} & Over a long period, hard to search within the storage space of an ML developer or a shared space of an ML team. & With database and ontological support, easy to query for all virtualized datasets with metadata-driven discovery. \\
    \midrule

    
     \raggedright\textbf{C$_1$ Effort duplication} & Commonly duplicated effort by different ML developers working on similar problems. & Standard queries enable reducing duplicated effort, and with search capability, reduce duplication across teams. \\
    \midrule
    
     \raggedright\textbf{C$_2$ Model maintenance} & Models are maintained by ML developers individually & Models are maintained centrally \\
    \midrule
    
     \raggedright\textbf{C$_3$ Quality and efficiency} & Varying efficiency and varying quality due to diverse programming skills among ML developers. & Efficient (and quality-assured) data operations provided centrally for all ML developers through APIs and UIs. \\
    \midrule
    
     \raggedright\textbf{C$_4$ Collaboration workflows} & Email-based coordination, informal code sharing, difficulty synchronizing work across team members, potential conflicts in concurrent data access. & Structured collaboration with shared workspaces, real-time collaboration features, conflict resolution mechanisms, and coordinated data access patterns. \\
    \midrule
    
     \raggedright\textbf{C$_5$ Knowledge sharing} & Limited or ad hoc sharing among ML developers & Systematic sharing among ML developers as well as teams \\
    \midrule
    
    
     \raggedright\textbf{D$_1$ Security and compliance} & Inconsistent security practices, difficulty enforcing data access policies, ad hoc compliance with institutional regulations, potential data leakage through uncontrolled copies. & Centralized security policies, role-based access control, automated compliance checking, and audit trails for all data operations. \\
    \midrule
        
     \raggedright\textbf{D$_2$ Documentation \& knowledge management} & Scattered documentation, knowledge silos, difficulty transferring expertise when developers leave, inconsistent documentation standards. & Centralized documentation with automated generation from configurations, searchable knowledge base, and standardized documentation templates. \\
    \bottomrule
    \\
    \multicolumn{3}{l}{Note: A: \emph{Core ML Operations}, B: \emph{Resource Management}, C: \emph{Development Efficiency}, and D: \emph{Operational Excellence.}}
\end{tabular}
\vspace{-8mm}
\end{table}

\section{Approach}
\label{sec:Approach}
In companies or organizations that have substantial ML activities, it is common to have a computational infrastructure to support many concurrent ML workflows. These ML activities are part of the long-term operations within such companies and organizations and involve one or more teams of ML developers and multiple other stakeholders. Hence, a technical solution to address the shortcomings in the middle column of Table \ref{tab:Comparison} is \emph{Data Virtualization}.   

Firstly, \emph{data virtualization} does not create explicit copies of the output datasets $\mathbf{Y}_1 \times \mathbf{Y}_1 \times \ldots \times \mathbf{Y}_{N_D}$. Instead, it creates the equivalent virtualized copy for each $\mathbf{Y}_j (j=1, 2, \ldots, N_D)$, together with the necessary links to the source datasets and the transformation functions. Hence, Eq.\,\ref{eq:Transformation} is rewritten as:
\begin{equation}\label{eq:Virtualization}
    V_k = \mathbf{X}_1 \times \ldots, \times \mathbf{X}_{N_S}
    \longrightarrow \bigl[ \Phi(F_k), \Psi(\mathbf{X}_1), \ldots, \Psi(\mathbf{X}_{N_S}) \bigr] \times
    \mathbf{Z}_1 \times \ldots \times \mathbf{Z}_{N_D} 
\end{equation}
where, $V_k$ is the virtualized version of the original data transformation $F_k$, $\Phi(F_k)$ is a linking identifier of (or path to) $F_k$, $\Psi(\mathbf{X})$ is a linking path to a source dataset $\mathbf{X}$, and $\mathbf{Z}_j$ is the virtualized version of $\mathbf{Z}_j$, $(j=1,2, \ldots, N_D)$.
With $\Phi(F_k)$ and $\Psi(\mathbf{X}_1), \ldots, \Psi(\mathbf{X}_{N_S})$, a virtualization service can dynamically retrieve or compute $\mathbf{Y}_j$ in place of $\mathbf{Z}_j$ $j=1,2,\ldots,N_D$ on demand.

In practice, most data virtualization processes are relatively simple, typically $N_S = N_D = 1$, and complex data virtualization can often be decomposed into a series of simple data transformations. However, the generalized notation in Eq.\,\ref{eq:Virtualization} can represent some complex data transformations, such as a feature specification based on multivariate data inputs from different data sources.

\begin{figure}[t]
    \centering
    \includegraphics[width=\linewidth]{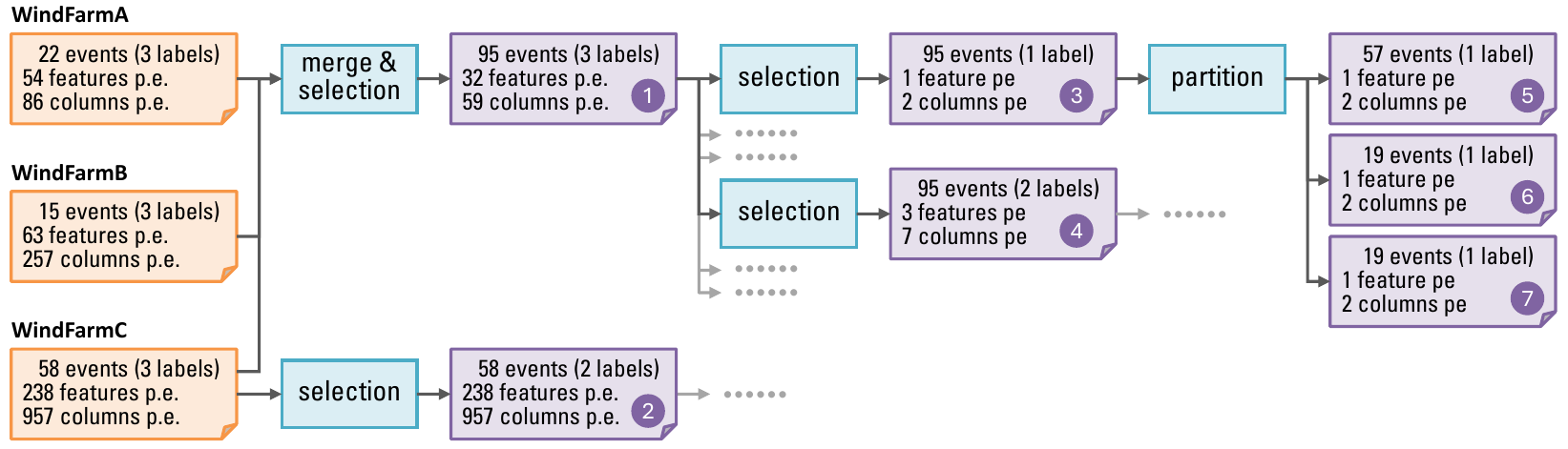}
    \caption{An example sequence of virtualization operations for transforming three original datasets (on the left) to three virtual datasets for training, validation and testing (on the right). The figure also shows other possible sequences of transformations, illustrating the need for reducing the storage space while maintaining data provenance.}
    \label{fig:V-Ops}
\end{figure}

Fig. \ref{fig:V-Ops} shows several examples of virtualization operations. The three wind farm datasets consist of wind turbine event records \cite{gueck2024scada}. These datasets are part of the publicly available CARE to Compare collection, which includes 89 years of SCADA time series data from 36 turbines located in three farms in Europe. They were designed to benchmark anomaly/fault detection methods in wind turbine operations. In the original off-line datasets, each event is stored as a .csv file, with data of different time stamps stored in rows and different features stored in columns. Each of these .csv files has about 50K$\sim$65K rows and up to 957 columns.
Because each farm uses a different set of sensors, the datasets have a different number of features. Some features are multivariate features, leading to different numbers of columns among the event records.

We can identify 59 columns that are common among the three datasets, select these shared columns, and create a merged dataset as indicated by \ballnumber{1} in Fig. \ref{fig:V-Ops}. This virtual dataset only consists of the IDs of 95 virtual data objects and their link paths to the original event records in the three datasets on the left. The virtual dataset has its own randomized IDs in case of duplicate identifiers among the original datasets.

The Wind Farm datasets were collected for training classification models with one label (\emph{normal} and \emph{anomaly}) and segmentation models with two labels (\emph{event start} and \emph{event end}). Different ML developers may create further virtual datasets, e.g., in the case of \ballnumber{3}, selecting one unary feature to train a classification model, and in the case of \ballnumber{4}, selecting three unary features and one 4-ary features to train a segmentation model. These virtual datasets can be further partitioned into training, validation, and testing datasets as indicated by \ballnumber{5}, \ballnumber{6}, and \ballnumber{7}.

One can easily notice that there are many options for selecting features in the cases of \ballnumber{3} and \ballnumber{4}. If one needs to experiment with different selections of features, one can create virtualized datasets with little space while maintaining data provenance.

Not only can such data virtualization services improve the cost-effectiveness in data storage, but they can also provide many benefits, e.g., maintaining data provenance, sharing data, sharing data transformation functions, maintaining consistency and comparability among experiments conducted by different ML developers, and so on. In Table \ref{tab:Comparison}, we list the benefits of an infrastructure with data virtualization, in contrast to the shortcomings arising from individually managed data transformations without data virtualization services. These shortcomings and benefits are grouped into four categories, i.e., A: \emph{Core ML Operations}, B: \emph{Resource Management}, C: \emph{Development Efficiency}, and D: \emph{Operational Excellence}.
   
\begin{figure}[t]
    \centering
    \includegraphics[width=1\linewidth]{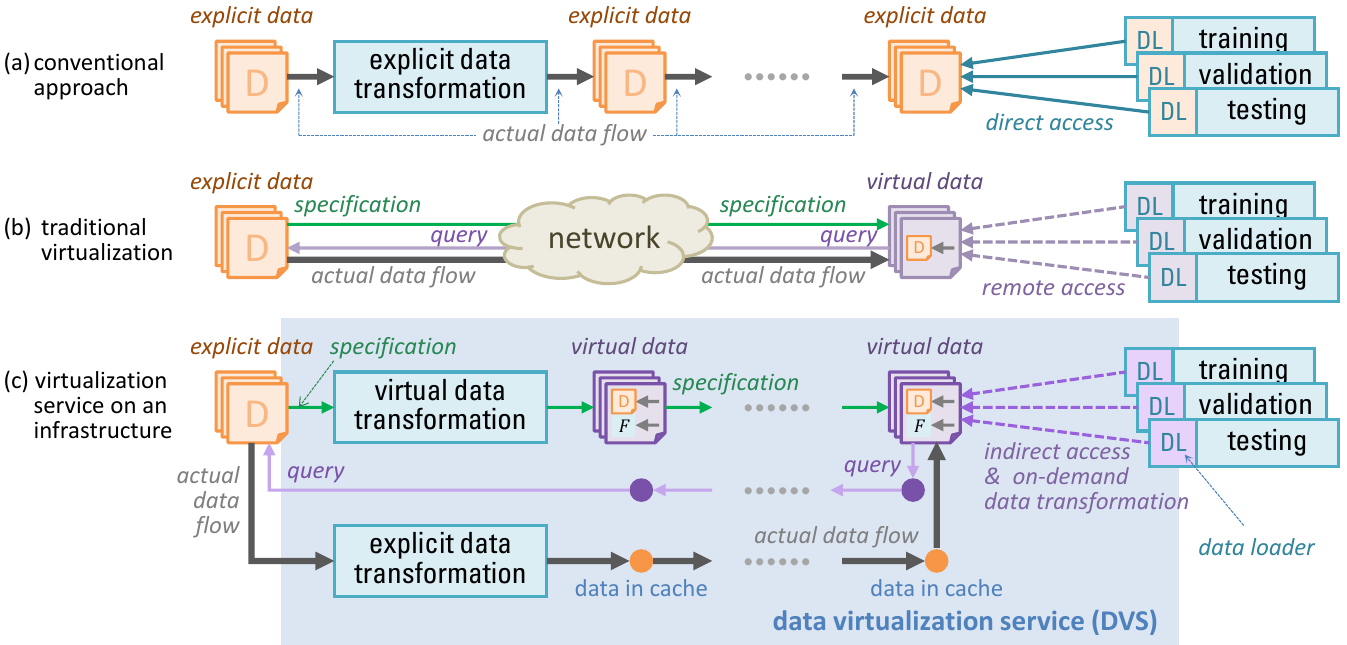}
    \caption{
    (a) A conventional data flow for data preparation prior to training, validation, and testing.
    (b) A traditional data virtualization workflow focusing on remote access across different infrastructures.
    (c) A virtualization workflow focusing on data transformations within an infrastructure.
    While this work focuses on (c), it also supports the hybrid data flows between (a) and (c), and can be extended to include (b).
    }
    \label{fig:workflow}
\end{figure}

Modern ML APIs (e.g., Pandas, Polars, PyTorch, NumPy, scikit-learn, etc.) provide robust APIs to support a wide range of data transformation operations, including the operations mentioned in the previous sections. It is not our intention to replace these APIs, but to utilize them in developing data virtualization services. As illustrated in Fig.~\ref{fig:workflow}(a), ML developers commonly use such APIs to implement a series of data transformations from source data to data that can be loaded into ML training, validation, and testing software. For individual ML developers, it is usually convenient to create explicit datasets in such workflows, allowing the ML software to load the aforementioned datasets directly via its data loader.

Traditionally, data virtualization focuses on remote access of data across different infrastructure, as illustrated in Fig.~\ref{fig:workflow}(b). A virtual dataset essentially contains only the necessary information to access remote data. Hence, creating a virtual dataset is an action of \emph{specification}, and accessing data is an action of query, followed by the actual data flow to transfer the data to the ML software data loader.

In this work, we extend the notion of data virtualization to include data transformations within an infrastructure. As shown in Fig.~\ref{fig:workflow}(c), the notion of specification now includes not only the path links to the proceeding data, which can be explicit or virtual, but also the links for calling data transformation functions. When a data loader in the ML software accesses a virtual dataset, it invokes a series of queries that trace back to one or more explicit datasets, followed by a series of data transformations to load the data into the ML software. Note that ML software may load an entire dataset or individually data objects in a dataset. The specifications, queues, and actual data transformations are all managed by the data virtualization service. Unlike the data flow in Fig.~\ref{fig:workflow}(a), the intermediate data is not stored as explicit data files, but as data in the cache to be accessed by the data transformation functions.
In the next section, we will detail the design and implementation of the data virtualization service.

\section{Design and Implementation}
\label{sec:design-implementation}

\begin{figure}[t]
    \centering
    \includegraphics[width=1\linewidth]{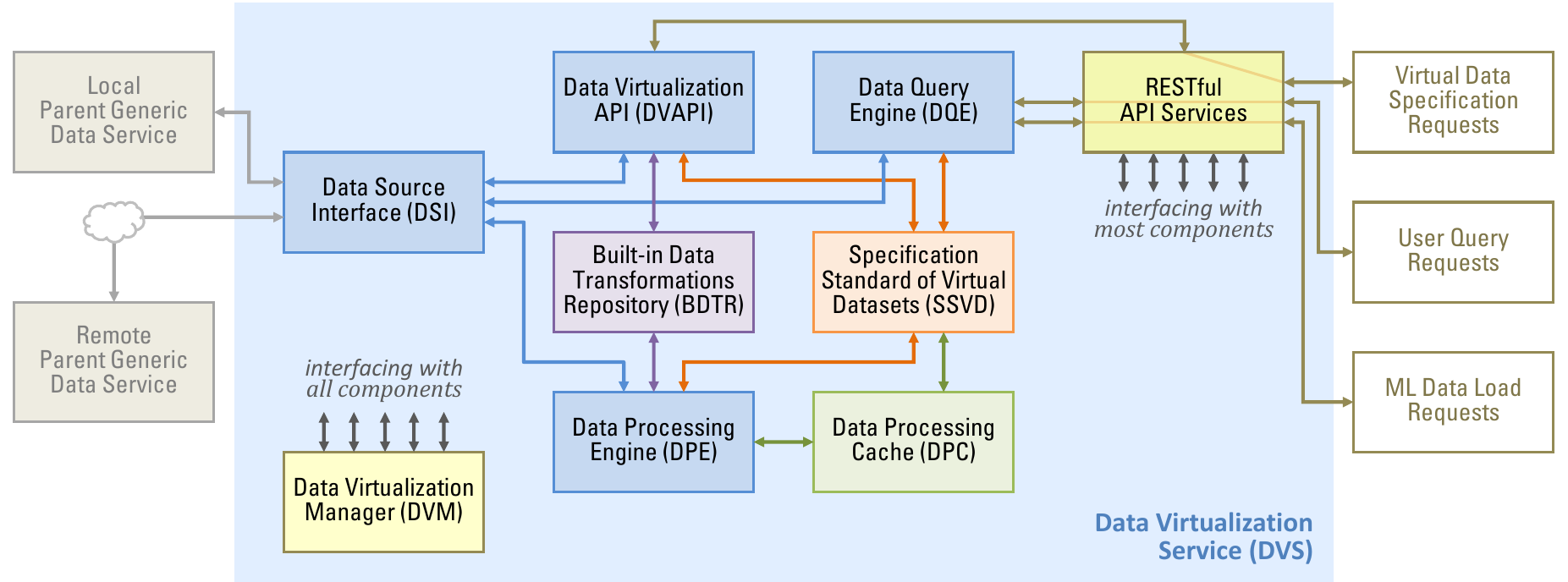}
    \caption{
    The service architecture of the data virtualization service and its components.
    }
    \label{fig:architecture}
\end{figure}

\textbf{Requirements.}
A \emph{data virtualization service} is part, or an add-on, of a general-purpose data service. Hence, its capability to support different storage systems (e.g., local server disks, databases, cloud object storage, remote data repositories, etc.), different data types (e.g., audio, videos, geographical data, etc.), file formats (e.g., CSV, NetCDF, PPN, etc.), and different ML APIs (e.g., Pandas, Polars, PyTorch, NumPy, scikit-learn, etc.),  depends on the ``parent'' general-purpose data service. The work reported here focused mainly on the development of a data virtualization service based on the existing capability of the ``parent'' data service, while we added capabilities to deal with some specific applications on demand.  

Our data virtualization service is designed to support \emph{ML model development} workflows. Currently, it is not intended for \emph{model deployment}, where dynamic data streams would be encountered. In the context of service development, a data virtualization service should meet the following system requirements:
\begin{itemize}
    \item
    Support all storage systems that the ``parent'' data service can support;
    \item
    Support data types and data formats that the ``parent'' data service can structurally understand (i.e., not stored as opaque data);
    \item
    Support all ML APIs available on the infrastructure, where the ``parent'' data service can access.
    \item
    Manage the relationships among explicit datasets and virtual datasets, allowing creation of new relationships (through specification of virtual data), and queries about existing relationships as well as navigation among them (e.g., during data loading, during users' search for existing datasets, etc.);
    \item
    Support a range of virtual data transformation functions as built-in virtualization operations, such as merge, selection, partition, and reorganization, together commonly-used functions for normalization and feature extraction. Note that there are numerous functions for the latter. We thus have an initial set of built-in virtualization operations, and gradually add more functions on demand.
    \item
    Provide a mechanism for advanced users to include user-defined data transformation programs as virtualization operations. 
\end{itemize}

\vspace{2mm}\noindent
\textbf{Design.}
As shown in Fig. \ref{fig:architecture}, the architecture of the \emph{Data Virtualization Service} (DVS) includes several system components:
\begin{itemize}
    \item
    \textbf{Data Virtualization Manager} (DVM) -- It maintains full knowledge of all current virtual datasets, as well as all explicit datasets managed by the ``parent'' data service for ML training, validation, and testing.
    It also enables the creation and removal of virtual datasets. It is the core component of DVS to meet the aforementioned requirements.
    \item
    \textbf{Data Source Interface} (DSI) -- This interface allows the DVS to interact with the ``parent'' data services for accessing explicit datasets managed by the ``parent'' services. All our current implementation involves only one ``parent'' data service, we are extending the current DSI to access another ``parent'' data service, which is located remotely and has a different service architecture.  
    \item
    \textbf{Specification Standard of Virtual Datasets} (SSVD) -- It provides a standard file format to specify virtual datasets. It enables a declarative specification of each virtual dataset in a way that DVM can understand, including details such as meta data, path links to source datasets, data hierarchy, selection criteria, call links to data transformations, and randomization settings. We adopted the structured YAML syntax for the SSVD.
    \item
    \textbf{Data Virtualization API} (DVAPI) -- It provides a programming interface to specify a virtual dataset as well as to query a virtual dataset.
    \item
    \textbf{Data Query Engine} (DQE) -- The query processing unit for processing a query from a data loader or from a user's query, and navigating path links backward (as shown in Fig. \ref{fig:workflow}(a) or forwards.   
    \item
    \textbf{Built-in Data Transformations Repository} (BDTR) -- It provides a list of data transformations, whose unique identifiers can be used by the DVAPI in creating a virtual dataset and by the data processing engine to perform the corresponding data transformations. Some data transformations require specialized implementation due to virtualization (e.g., merge, selection, partition, and reorganization), and others are built on top of existing ML APIs (e.g., normalization and feature extraction).
    \item
    \textbf{Data Processing Engine} (DPE) -- The computational unit to invoke data transformation functions during data loading. 
    \item
    \textbf{Data Processing Cache} (DPC) -- The DVS maintains a cache space for the DPE to process various data transformations along the dark gray path shown in Fig. \ref{fig:workflow}(c).
    \item
    \textbf{RESTful API Services} -- To maintain the transitions of service states of different activities in DVS (e.g., specifications, queries, data transformation, etc.)~\cite{Khan2022:TSC}.
\end{itemize}

\vspace{2mm}\noindent
\textbf{Implementation.}
We implemented our DVS primarily in Python (version 3.12). The REST API backend was implemented using FastAPI for high-performance, asynchronous request handling, with Pydantic for robust data validation and serialization.
Several data transformation functions were built on Pandas and NumPy.
We utilized DuckDB for efficient caching and analytical queries with columnar storage and PyArrow for optimized data interchange, along with other Python packages.
Although the software for the DVS was developed on a Windows laptop with Linux (WSL) via Docker, it was tested and deployed in a cloud infrastructure in the UK STFC (Science and Technology Facilities Council). The Windows laptop used for the development is equipped with 32 GB RAM and a 14-core Intel i7-13700H processor (20 threads with hyper-threading).

Compared to the conventional approach shown in Fig. \ref{fig:workflow}(a), the overhead of the DVS is mostly unnoticeable on the laptop in terms of speed. The speed difference on the STFC cloud infrastructure is negligible.  
We anticipate that the speed of users' queries for some virtual datasets will be affected slightly when the total number of virtual datasets increases, while the access queries (e.g., the purple line in Fig. \ref{fig:workflow}(c)) require only a trivial amount of time to track back to the source datasets. Meanwhile, the actual data flow (i.e., the dark gray line in Fig. \ref{fig:workflow}(c)) is faster than the data flow in Fig. \ref{fig:workflow}(a)) due to the provision of the data processing cache (DPC). When we take the time cost for individual ML developers to code and run the data transformation programs in Fig. \ref{fig:workflow}(a), Fig. \ref{fig:workflow}(c) represents substantial cost savings.


\section{Applications}
\label{sec:Application}
The data virtualization service described in the previous section has been used to support a number of ML workflows in several applications. As illustrated in Fig.~\ref{fig:Applications}, these applications include:

\begin{figure}[t]
\centering
\begin{tabular}{@{}c@{\hspace{4mm}}c@{\hspace{4mm}}c@{}}
    \includegraphics[width=50mm, height=35mm]{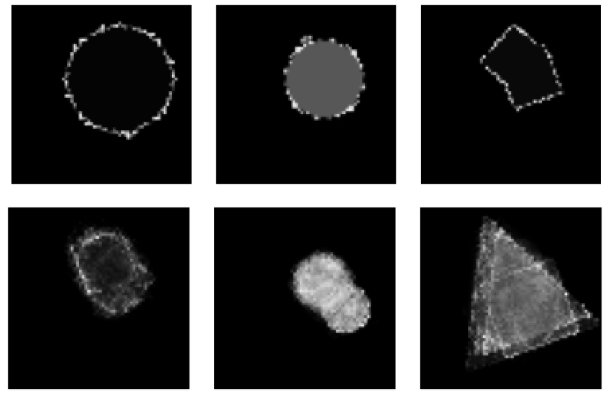} &
    \includegraphics[width=50mm, height=35mm]{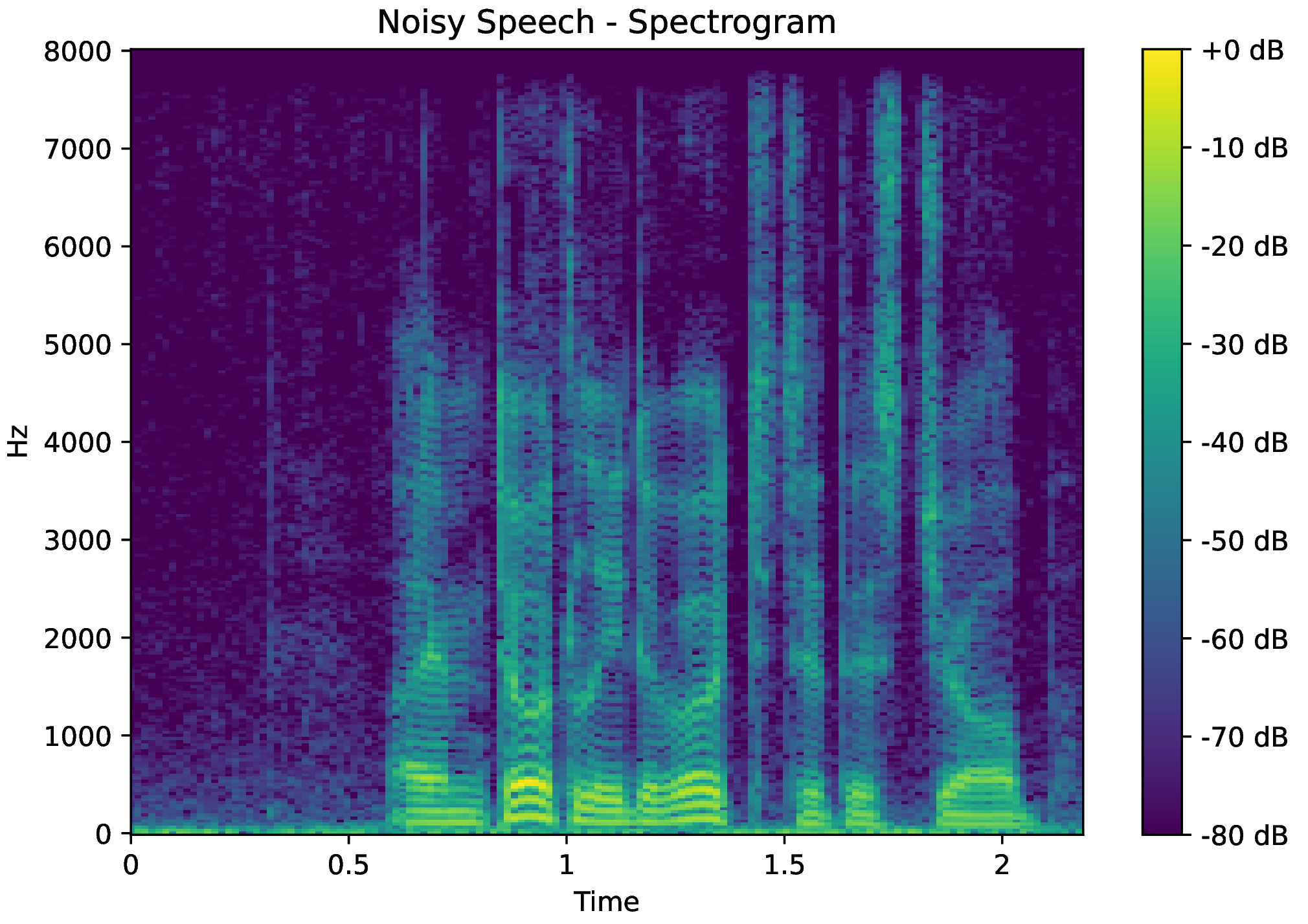} &
    \includegraphics[width=50mm, height=35mm]{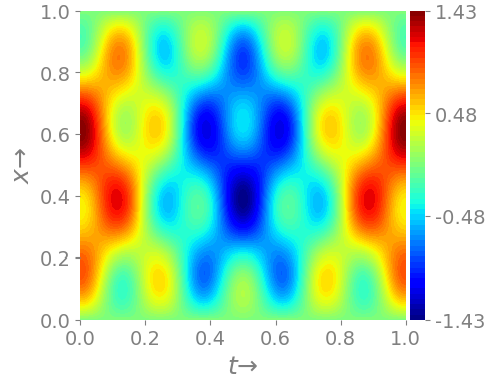} \\
    \scriptsize{(a) Images} &
    \scriptsize{(b) Speech signals} &
    \scriptsize{(c) Wave equations} \\[2mm]
    \includegraphics[width=50mm, height=35mm]{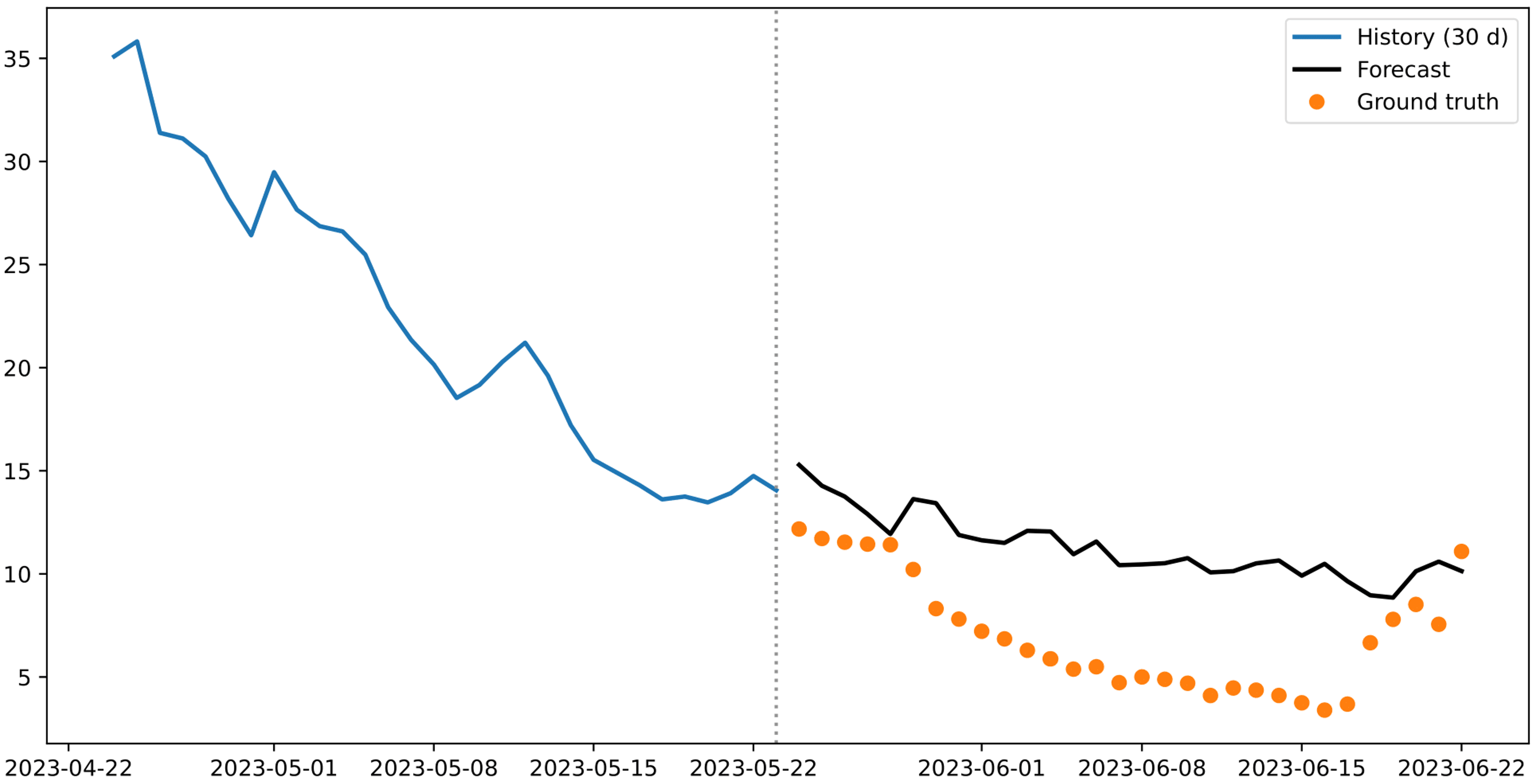} & 
    \includegraphics[width=50mm, height=35mm]{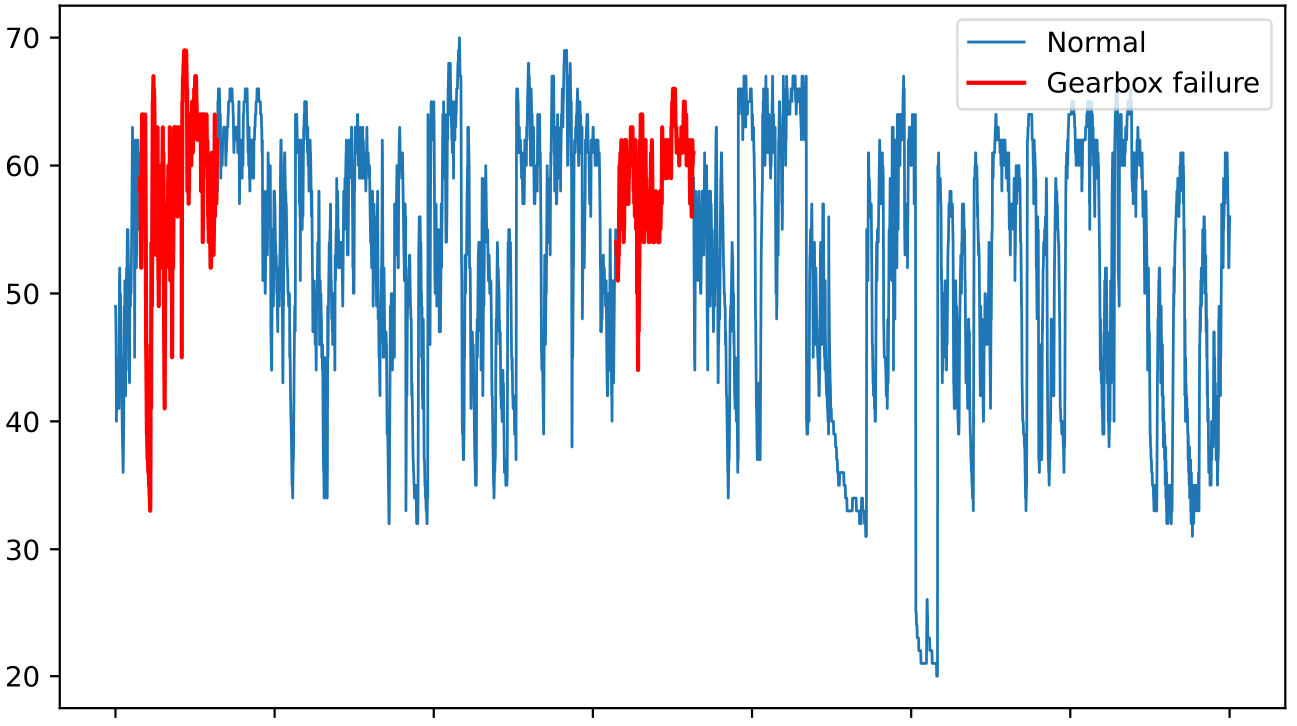} &
    \includegraphics[width=50mm, height=35mm]{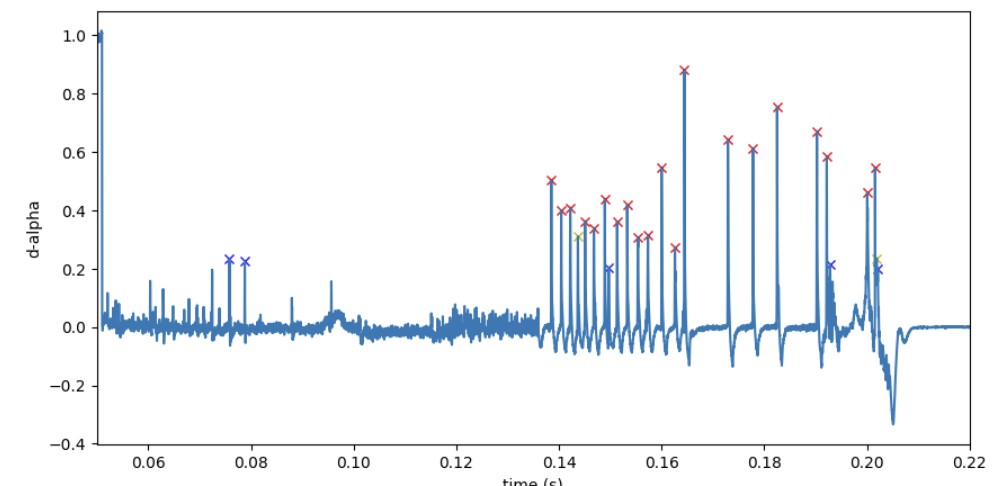} \\
    \scriptsize{(d) Time series (water level)} &
    \scriptsize{(e) Time series (wind turbine sensor)} &
    \scriptsize{(f) Time series (fusion diagnostics data)}
\end{tabular}
\caption{The data virtualization service currently supports six ML applications. There are multiple ML workflows in each application.}
\label{fig:Applications}
\end{figure}

\vspace{2mm}\noindent
\textbf{Embedding in Image Reconstruction and Classification.} This encompasses a number of theoretical studies, each corresponding to an ML workflow. The studies focus on the understanding of the functionality of embedding, in training different ML models. For example, a pivotal model is used for image reconstruction.    
A deep autoencoder employs an encoder (2048\,$\to$\,1024\,$\to$\,512\,$\to$\,256 bottleneck) to compress dense image input into low to high-dimensional latent embeddings (e.g., 256 dimensions), which are subsequently used to reconstruct image input using a symmetrically mirrored decoder (Fig. \ref{fig:Applications}(a)). This model achieves a mean similarity index of 0.93 in six geometric and visual categories, and its reconstruction fidelity is qualitatively assessed using side-by-side visual grids of original and reconstructed inputs for quality assurance.
As studies on this reconstruction model include experimentation with different geometric and visual attributes, many imagery datasets would need to be generated by a scene configuration program. The data virtualization service allows ML researchers to create datasets with virtual data objects containing only image attributes. Such virtual datasets can be processed and partitioned in the same way as normal datasets.   

In addition to this reconstruction model, there are workflows for training classifiers that use embeddings as an intermediate input. To train a classifier for each geometric category, one needs at least one ML workflow. Because the original imagery data objects are associated with multivariate data labels (e.g., shape, size, color, etc.), the data virtualization service can help select data objects for each sub-label and partition data to ensure the training, validation, and testing datasets with balanced sub-label distribution.

\vspace{2mm}\noindent
\textbf{Speech Enhancement.} Another application of our data virtualization service is speech enhancement using a UNet style convolutional neural network that takes speech spectrogram images as input and is trained with paired noisy/clean examples to perform denoising (Fig. \ref{fig:Applications}(b)). The encoder consists of multiple convolutional layers that down sample the time frequency map halving its dimensions at each stage, until a bottleneck of 128 feature channels is reached, using depthwise separable convolutions to limit parameter count. A corresponding up sampling decoder then reconstructs the denoised spectrogram, which we convert back to waveform using the inverse Short-time Fourier transform (STFT) (with 50\% overlap‑add). In standard evaluation sets, this model improves the average PESQ from 1.97 to 2.80 and yields a +12 dB gain in segmental SNR. Due to its small model size and per‑frame streaming operation (e.g., 25 ms latency), it can be seamlessly integrated into Voice over IP systems for robust, real‑time audio enhancement.

In addition, there are workflows for studying model pruning. The effectiveness of pruning is influenced by many interrelated factors, including the input data space, the model task, the network architecture, and the ML hyperparameters. Understanding the impact of these factors requires large-scale, multivariate experiments.
For example, the effectiveness of pruning is sensitive to noise and needs to be evaluated using data with different types and levels of noise augmentation. Since noise augmentation is a form of data transformation, experimentation on different types and levels of noise augmentation can benefit from data virtualization to avoid the storage of many versions of noise augmented speech data, while maintaining the provenance of the experiments.

\vspace{2mm}\noindent
\textbf{River Flow/Water Level Forecasting.} One large application of our data virtualization service is the ML development related to
river and flood management. In the UK, there are some 1,500 river gauging stations and some 1,000 rainfall monitoring stations measuring  river flow and rainfall, respectively. Such historical data is used to develop ML models to forecast volumetric flow rate and water level at each gauging station. There are many ML workflows in this application, some for training forecasting models with only station-specific data, some for training models that take the topology of river networks into account, some for training models with both river data and rainfall data, and so on. 
For example, one workflow focuses on a dual-encoder transformer model that ingests past flow and rainfall records from station $S_i$ ($i=1,…,400$), along with exogenous meteorological input, to predict river-flow levels for the next 30 days (Fig. \ref{fig:Applications}(d)). This model achieves a mean absolute error (MAE) of 1.28 and a root mean squared error (RMSE) of 1.37.

Data virtualization plays a critical role in supporting the diverse range of ML workflows, as the same data collected by a river gauging station or rainfall gauge is used in many different workflows, often with different data transformations.

\vspace{2mm}\noindent
\textbf{Wind Turbine SCADA Data For Early Fault Detection.}
There are three datasets that are collected in different ways. In Fig. \ref{fig:V-Ops}, we illustrate three types of data transformation, i.e., selection, merge, and partition. Similarily, there are many workflows in this application, which are primarily divided into two groups. One group includes a large number of workflows for training different sub-models using the time series data of individual sensors. An ensemble model is then derived from the trained sub-models. The ensemble model and the individual sub-models are all classifiers trained using XGBoost. The ensemble models identify impending gearbox faults with 93\,\% accuracy (F1 = 0.91) across healthy operational states and diverse gearbox failure modes (Fig. \ref{fig:Applications}(e)).
The second group includes workflows for training models to detect the starting and ending points of each fault in time series that are labelled ``anomaly''.

\vspace{2mm}\noindent
\textbf{Partial Differential Equations (PDE)}.
There have been growing research activities in developing ML models for approximating numerical PDE solvers. For example, Fig.~\ref{fig:Applications}(c) shows the ground truth solution to solve a set of wave equations. Physics-informed machine learning models~\cite{Karniadakis2021:NRP} are trained with spatial or spatiotemporal coordinates randomly sampled from such ground truth data.
Physical constraints are incorporated into the loss function used in the training process, steering the learning towards the ground truth solution from scattered data points. 
However, the effectiveness of training critically depends on the strategies used for data sampling, while the effectiveness of sampling strategies can sometimes be PDE-dependent. 
 
Traditionally, ML developers manually generate, store and manage thousands of training point configurations, leading to significant storage overhead or difficulties in reproducing successful experiments.
They explore multiple sampling strategies systematically, e.g., uniform grid sampling, adaptive sampling near boundaries, Latin hypercube sampling for better space coverage, etc. 
Furthermore, each strategy requires testing across various point densities (from 1,000 to 100,000 points), different spatial distributions, and multiple random seeds to ensure robustness.
With data virtualization, the sampling strategies are implemented as different data transformation functions to select scattered data points from the ground truth data. This enables rapid and reliable data generation in the experimental workflows for training ML PDE models, accelerating the model development process, and maintaining the critical provenance information (e.g., strategy, random seed, etc.) for the repeatability of the experiments.

\vspace{2mm}\noindent
\textbf{Fusion Data Analysis.} 
Another application of data virtualization is the analysis of time series diagnostic data from the Mega Ampere Spherical Tokamak (MAST) \cite{Jackson2024:MAST} reactor at UKAEA (Fig.~\ref{fig:Applications}(d)). The MAST archive comprises experimental discharges ``shots'' in the order of about 40,000, each capturing high frequency (\textasciitilde500Hz) signals with up to 50,000 sampling points per signal. Examples of such signals are deuterium emission $D_{alpha}$, plasma current ($I_p$), plasma density $\eta_e(r)$, neutral beam power ($P_{NBI}$), magnetic field on plasma axis ($B_T$), and so on.
Making sense of cross-signal relationships is the key to understanding plasma behavior. It enables accurate detection of plasma instability, such as detection of edge localized modes (ELMs), and thereby facilitates the optimal operation of fusion reactors. To analyze cross-signal relationships, ML developers select multiple signals across shots to identify a subset of relevant signals for each specific task, i.e., confinement mode classification.
The data virtualization service allows seamless access to different signals without the need for physical data consolidation, improving the cost-effectiveness of these ML workflows in terms of data storage, programming effort, data provenance, and team work.  


\section{Conclusion}
\label{sec:conclusion}

In this paper, we have extended the concept of data virtualization for creating and managing virtual datasets used by many machine learning (ML) workflows. These virtual datasets stores not only path links to proceeding datasets, but also call links to data transformation functions. We presented our design of a data virtualization architecture and reported its implementation and six application case studies. While the current data virtualization service is built on top a local generic data service where explicit datasets are stored, we are developing a service interface with a remote infrastructure that has a different type of generic data service. We also plan to make the software for our data virtualization service available in the public domain as open-source software, and are working on the necessary system documentation.

Our approach provides a cost-effective replacement for the conventional practice, in which individual ML developers or individual teams develop and run their own data transformation programs.
It enables ML developers to focus on experimental designs rather than data wrangling tasks for supporting different experiments, facilitates seamless data provenance management rather than burdening ML development with rigorous record keeping.
It brings about storage efficiency through replacing explicit datasets with virtual datasets rather than through ad hoc removal of used datasets.
The six applications currently supported by the data virtualization service demonstrate that with such a service, ML developers are more enthusiastic and more creative in exploring different ML methods and experiments when working on the same datasets.

In addition to aforementioned interface with a remote infrastructure, we are working on extending the Built-in Data Transformations Repository (BDTR), and monioring the scalability of the data virtualization service when the numbers of applications and ML workflows increase in the coming years. We will also introduce data visualization capability for visualizing application specific data provenance as well as holistic data ontology managed by the data virtualization service.


\section*{Acknowledgment}
This project was funded by UKRI/EPSRC grant EP/X029557/1.

\keywords{Data Virtualization \and Machine Learning \and Services \and Service Oriented Architecture \and SOA \and Service Operations}

\bibliographystyle{unsrt}  
\bibliography{bibliography}

\begin{thebibliography}{10}

\bibitem{halevy2005}
Alon~Y. Halevy, Naveen Ashish, Dina Bitton, Michael Carey, Denise Draper, Jeff Pollock, Arnon Rosenthal, and Vishal Sikka.
\newblock {Enterprise Information Integration: Successes, Challenges and Controversies}.
\newblock In {\em ACM SIGMOD}, pages 778--787, 2005.

\bibitem{gu2024}
Zhenzhen Gu, Francesco Corcoglioniti, Davide Lanti, Alessandro Mosca, Guohui Xiao, Jing Xiong, and Diego Calvanese.
\newblock {A Systematic Overview of Data Federation Systems}.
\newblock {\em Semantic Web}, 15(1):107--165, 2024.

\bibitem{gadepallyBigDAWG2017}
Vijay Gadepally, Kyle O'Brien, Adam Dziedzic, Aaron Elmore, Jeremy Kepner, Samuel Madden, Tim Mattson, Jennie Rogers, Zuohao She, and Michael Stonebraker.
\newblock {BigDAWG Version 0.1}.
\newblock In {\em 2017 {{IEEE High Performance Extreme Computing Conference}} ({{HPEC}})}, pages 1--7, 2017.

\bibitem{xiao2019}
Guohui Xiao, Linfang Ding, Benjamin Cogrel, and Diego Calvanese.
\newblock {Virtual Knowledge Graphs: An Overview of Systems and Use Cases}.
\newblock {\em Data Intelligence}, 1(3):201--223, 2019.

\bibitem{Lin2023}
Weiwei Lin, Chennian Xiong, Wentai Wu, Fang Shi, Keqin Li, and Minxian Xu.
\newblock {Performance Interference of Virtual Machines: A Survey}.
\newblock {\em ACM Computing Surveys}, 55:1--37, 12 2023.

\bibitem{lans2022}
Rick F. van~der Lans.
\newblock Data virtualization in the time of big data.
\newblock White paper, 2022.

\bibitem{muniswamaiah2019}
Manoj Muniswamaiah, Tilak Agerwala, and Charles Tappert.
\newblock {Data Virtualization for Decision Making in Big Data}.
\newblock 10(5):45--53.

\bibitem{jarwar2019}
Muhammad~Aslam Jarwar, Sajjad Ali, and Ilyoung Chong.
\newblock {Microservices Model to Enhance the Availability of Data for Buildings Energy Efficiency Management Services}.
\newblock {\em Energies}, 12(3):360, 2019.

\bibitem{balakrishnan2024}
Anandaganesh Balakrishnan.
\newblock {Data Virtualization Architecture Framework Using Multi-Engine Data Platforms for Big Data Analytics and Machine Learning}.
\newblock {\em International Journal of Computer Trends and Technology}, 72(2):82--91, 2024.

\bibitem{Kennedy2023}
Jason Kennedy, Vishal Sharma, Blesson Varghese, and Carlos Reaño.
\newblock {Multi-Tier GPU Virtualization for Deep Learning in Cloud-Edge Systems}.
\newblock {\em IEEE Transactions on Parallel and Distributed Systems}, 34:2107--2123, 7 2023.

\bibitem{Zhang2020}
Jin Zhang, Zhuocheng Ding, Yubin Chen, Xingguo Jia, Boshi Yu, Zhengwei Qi, and Haibing Guan.
\newblock {GiantVM: a type-II hypervisor implementing many-to-one virtualization}.
\newblock In {\em Proceedings of the 16th ACM SIGPLAN/SIGOPS International Conference on Virtual Execution Environments}, pages 30--44. ACM, 3 2020.

\bibitem{Chatziantoniou2021}
Damianos Chatziantoniou and Verena Kantere.
\newblock {DataMingler: A Novel Approach to Data Virtualization}.
\newblock In {\em Proceedings of the 2021 International Conference on Management of Data}, pages 2681--2685. ACM, 6 2021.

\bibitem{Scheiderer2020}
Christian Scheiderer, Timo Thun, Christian Idzik, Andrés~Felipe Posada-Moreno, Alexander Krämer, Johannes Lohmar, Gerhard Hirt, and Tobias Meisen.
\newblock {Simulation-as-a-Service for Reinforcement Learning Applications by Example of Heavy Plate Rolling Processes}.
\newblock {\em Procedia Manufacturing}, 51:897--903, 2020.

\bibitem{gueck2024scada}
C.~G{\"u}ck and C.~Roelofs.
\newblock {Wind Turbine SCADA Data For Early Fault Detection}, 2024.

\bibitem{Khan2022:TSC}
Saiful Khan, Phong~Hai Nguyen, Alfie Abdul-Rahman, Euan Freeman, Cagatay Turkay, and Min Chen.
\newblock {Rapid Development of a Data Visualization Service in an Emergency Response}.
\newblock {\em IEEE Transactions on Services Computing}, 15(3):1251--1264, 2022.

\bibitem{Karniadakis2021:NRP}
George~Em Karniadakis, Ioannis~G Kevrekidis, Lu~Lu, Paris Perdikaris, Sifan Wang, and Liu Yang.
\newblock {Physics-informed machine learning}.
\newblock {\em Nature Reviews Physics}, 3(6):422--440, 2021.

\bibitem{Jackson2024:MAST}
Samuel Jackson, Saiful Khan, Nathan Cummings, James Hodson, Shaunde Witt, Stanislas Pamela, Rob Akers, and Jeyan Thiyagalingam.
\newblock {FAIR-MAST: A Fusion Device Data Management System}.
\newblock {\em SoftwareX}, 27:101869, 2024.

\end{thebibliography}

\end{document}